\documentclass[]{aastex}
\usepackage{emulateapj5}

\newcommand{\etal}{{\it et al.~}}

\newcommand{\arcm}{{$^\prime\,$}}
\newcommand{\arcs}{{$^{\prime\prime}\,$}}    
\newcommand{\z}{{z^\prime}}    

\makeatletter

\begin{document}


\title{Weak Lensing Discovery and Tomography of a Cluster at $z=0.68$}

\shorttitle{Weak Lensing Discovery of a Cluster at $z=0.68$}
 
\author{D. Wittman\altaffilmark{1}, V. E. Margoniner\altaffilmark{1},
J.~A. Tyson\altaffilmark{1}, J.~G. Cohen\altaffilmark{2}
A.~C. Becker\altaffilmark{1}, I.~P. Dell'Antonio\altaffilmark{3}}
      
\altaffiltext{1}{Bell Laboratories, Lucent Technologies, Murray Hill,
NJ 07974; wittman,vem,tyson,acbecker@science.lucent.com}
\altaffiltext{2}{Caltech, Pasadena, CA 91125; jlc@astro.caltech.edu}
\altaffiltext{3}{Physics Department, Brown University, Providence, RI
02912; ian@het.brown.edu}
 
\begin{abstract}

We report the weak lensing discovery, spectroscopic confirmation, and
weak lensing tomography of a massive cluster of galaxies at $z=0.68$,
demonstrating that shear selection of clusters
works at redshifts high enough to be cosmologically interesting.  The
mass estimate from weak lensing, $11.1 \pm 2.8 \times 10^{14}\
(r/Mpc)\ M_\odot$ within projected radius $r$, agrees with that
derived from the spectroscopy ($\sigma_v = 980$ km s$^{-1}$), and with
the position of an arc which is likely to be a strongly lensed
background galaxy.  The redshift estimate from weak lensing tomography
is consistent with the spectroscopy, demonstrating the feasibility of
baryon-unbiased mass surveys.  This tomographic technique will be able
to roughly identify the redshifts of any dark clusters which may
appear in shear-selected samples, up to $z\sim 1$.
\end{abstract} 


\section{Introduction}

Galaxy clusters are important tools for studying the formation of
structure over cosmic time and for probing cosmological parameters
(Haiman, Mohr \& Holder 2001).  Such studies depend crucially on
the selection of unbiased samples of clusters covering a broad mass and
redshift range.  (``Cluster'' here indicates any mass concentration,
regardless of galaxy or gas content, as mass is the most important
parameter for cosmological tests.)  The most well-established
selection techniques are based on emission of visible-wavelength light
from member galaxies (Gladders \& Yee 2000 and references therein) or
of X-rays from hot intracluster gas (Borgani \& Guzzo 2001 and
references therein).  A newer technique uses the Sunyaev-Zel'dovich
effect, in which the cosmic microwave background is modified in its
passage through the intracluster medium (Carlstrom, Holder \& Reese
2002 and references therein).  Each of these methods depends on the
presence of baryons and on other physical conditions within the
cluster and therefore may introduce some bias.

In contrast, weak lensing (see Bartelmann \& Schneider 2001 for a
review) has the potential to select clusters independent of their
baryon content, dynamical state, and star formation history (Schneider
1996).  However, surveying for clusters via shear selection is still
in its infancy.  The only shear-selected mass with a spectroscopic
redshift is at $z=0.27$ (Wittman \etal 2001, hereafter W2001), whereas
cosmological effects on cluster abundances are expected to become
significant only well above this redshift.  More recently, Dahle \etal
(2003) and Schirmer \etal (2003) each identified several
shear-selected masses with redshifts determined from two-color
photometry as $z \sim 0.5$.  In other cases, (Erben \etal 2000; Umetsu
\& Futamase 2000; Clowe, Trentham \& Tonry 2001; Miralles \etal 2002),
the object causing the shear has not been assigned a redshift.
Weinberg \& Kamionkowski (2003) calculate that up to 20\% of clusters
in shear-selected surveys are expected to be optically dark. However,
without a redshift, the masses of these ``dark clusters'' cannot be
computed.  Hence mass-to-light ratios (or even limits) cannot be
computed either, and it is unclear just how dark these clusters are.
Nonspectroscopic means of determining their redshifts (and therefore
masses and derived parameters) must be developed.

Here we report the discovery of a shear-selected cluster at $z=0.68$,
which demonstrates that this technique can cover a significant
redshift range.  As in W2001, we also derive a lens redshift from weak
lensing tomography, by fitting the relation between shear and source
photometric redshift.  This redshift agrees with the spectroscopic
value, but is derived entirely independently.  This is the first
demonstration that even high-redshift ``dark'' clusters could be
assigned redshifts, thus making them cosmologically useful.  The
observations were obtained as part of the Deep Lens Survey (DLS) and
cover only a few percent of its area.  When complete, the survey will
yield a sample of $\sim 200$ shear-selected clusters.

\section{Imaging and Photometric Redshifts from the Deep Lens Survey}

The DLS (Wittman \etal 2002; see also http://dls.bell-labs.com) is an
ongoing deep $BVR\z$ imaging survey of six 2$^\circ \times 2^\circ$
fields using the Mosaic imagers on the KPNO and CTIO 4-m telescopes.
With total exposure times in $BVR\z$ of 12, 12, 18, and 12 ksec
respectively, it will reach a depth of 29, 29, 29, and 28 mag
arcsec$^{-2}$.  The $R$ filter is used when the seeing is 0.9\arcs\ or
better, to optimize its utility for lensing studies.  All shape
measurements are done in $R$, where the enforced good seeing will
provide a shear-selected survey which is more uniform and more
sensitive than would be possible with typical atmospheric conditions.
The other filters provide color information for photometric redshift
estimates, and are observed in when the seeing is worse than 0.9\arcs.
Photometric calibration is provided by observations of standard star
fields, using the calibrated catalog of Landolt (1992) for $BVR$ and
the most recent calibrated $\z$ catalog from D. Tucker (private
communication).

The Mosaic cameras provide 8k $\times$ 8k pixels subtending
0.257\arcs\ each, for a 35\arcm\ square field.  Each 35\arcm\
``subfield'' of the survey is imaged with 20 dithered exposures in each
filter, with dithers up to 200\arcs\ to provide good flatfielding.
Adjacent subfields will be stitched together into full 2$^\circ$
fields after all the contiguous data have been acquired, but for now
coaddition and analysis takes place on a subfield-by-subfield basis.
For this paper we are considering one particular subfield which has
been completed, centered at 10:54:43 -05:00:00 (J2000).  Several other
subfields have been completed, and shear-selected clusters tentatively
identified, but this cluster is the first to receive spectroscopic
confirmation.  This is due to the presence of the likely strongly
lensed arc, which gave us the confidence to arrange for spectroscopy
even before the weak lensing analysis was completed.  Therefore, this
cluster may not be typical of the final DLS shear-selected sample.
However, the characteristics of the data, such as number of sources
per square arcminute and photometric redshift accuracy, are
typical.

We observed this field with the CTIO 4-m Blanco telescope in 2000,
2001, and 2002 as part of the DLS imaging campaign.  Full details of
the data processing are given in Wittman \etal (2002), but we
summarize here.  We processed the data through flatfielding with
standard tasks from the IRAF package {\it mscred}, then registered and
combined them with custom software.

Before combining the $R$ images (the only bandpass intended for
lensing analysis), we first correct each exposure for point-spread
function (PSF) anisotropy using the procedure of Fischer \& Tyson
(1997).  This is necessary for each exposure because some observing
conditions, such as focus and guiding errors, change from exposure to
exposure.  Even within an exposure, the procedure is applied
separately for each CCD, in case there are piston differences between
the CCDs.  Briefly, the procedure is to find stars based on their
locus in the magnitude-size diagram; derive a fit to the spatial
variation of the PSF moments, clipping outliers which tend to be
interloping galaxies; and convolve the image with a kernel which, at
each point, is aligned orthogonal to the interpolated PSF at that
point.  Each CCD typically has 50--100 unsaturated, unambiguous stars,
which provide for a second-order polynomial fit to the spatial
variation.  We also apply this procedure to the combined $R$ image, to
reduce the effects of any small registration errors.  In this case, we
find $\sim 1000$ stars and use a fourth-order polynomial fit.  The
full-width at half-maximum (FWHM) of the $R$ image, after all combines
and convolutions, is 0.96 arcsec.  For comparison, the FWHM of the
unconvolved $B$, $V$, and $\z$ images are 1.03, 0.96, and 1.39 arcsec
respectively.

To produce photometric redshifts, we make matched-isophote catalogs
with detection in $R$ band using SExtractor (Bertin \& Arnouts 1996),
and use the photometry as input to a modified version of the HyperZ
photometric redshift package (Bolzonella, Miralles \& Pell\'{o} 2000).
The modification is an important one for the DLS.  Because of the
limited filter set, there are some color degeneracies.  That is, the
observed colors of a galaxy may be as well matched to one template at
low-redshift as to another template at high redshift.  Therefore we
add a luminosity function prior, as described in W2001, which
generally resolves the ambiguity (see Benitez 2000 for detailed
examples).  The DLS filter set differs from that of W2001, so we used
different luminosity function parameters, $M_R^* = -22.0$ and $\alpha
= -1.24$.  We adopt a cosmology in which $H_0 = 70$, $\Omega_m = 0.3$,
and $\Omega_\Lambda = 0.7$ throughout this work.

We verified the accuracy of the photometric redshifts in this subfield
by comparison with spectroscopic redshifts of 22 galaxies in the range
$0.36 < z_{\rm spec} < 0.98$ obtained at Keck (see
Section~\ref{sec-keck}), plus 49 galaxies in the range $0.04 < z_{\rm
spec} < 0.36$ from the 2dFGRS public data (Colless \etal\ 2001).  Thus
the full spectroscopic sample for this 35\arcm\ subfield contains 71
galaxies in the range $0.04 < z_{\rm spec} < 0.98$.  Like many other
authors, we measure the difference between photometric and
spectroscopic redshifts in terms of the quantity $\delta z = (z_{\rm
spec} - z_{\rm phot}) / (1 + z_{\rm spec})$, which is just the
percentage error in the quantity $1+z_{\rm spec}$.  This encodes the
fact that a redshift error of a given size is more important at low
redshift than at high redshift.

\begin{figure}
\epsscale{1.0}
\plotone{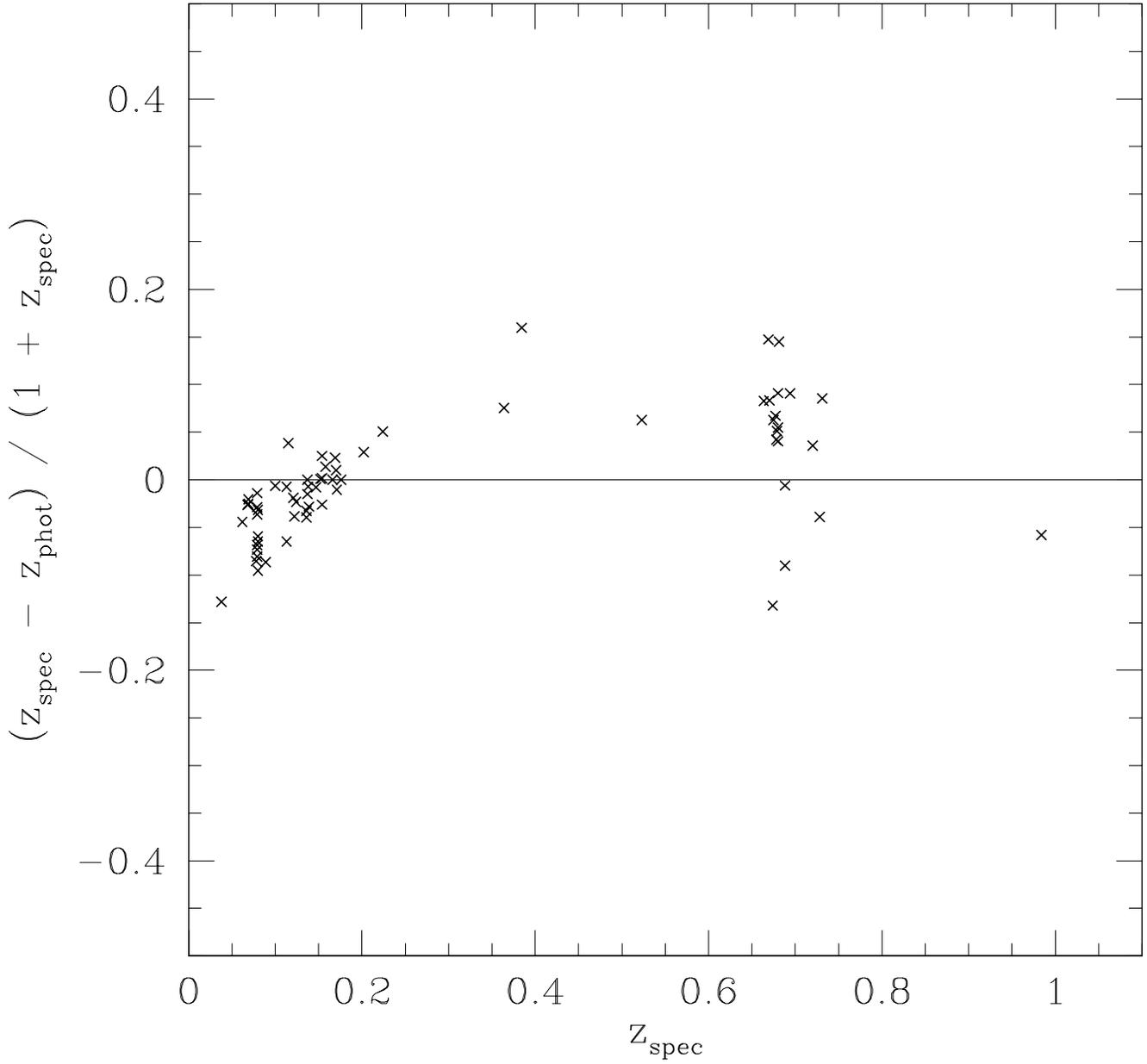}  
\caption{ The quantity $\delta z = (z_{\rm spec} - z_{\rm phot}) / (1
+ z_{\rm spec})$ versus spectroscopic redshift, for the 71 galaxies
with spectra.  The rms value of $\delta z$ is 0.065, and the range is
$-0.16 < \delta z < 0.16$.  This per-galaxy accuracy is sufficient for
lensing work, because it is significantly less than the inherent shape
noise in each galaxy.  The mean $\delta z$ averaged over all redshifts
is vanishingly small (-0.0014).
\label{fig-zphot}}
\end{figure}

Figure~\ref{fig-zphot} shows a scatterplot of $\delta z$ versus
$z_{\rm spec}$.  The rms value of $\delta z$ is 0.065, and the range is
$-0.16 < \delta z < 0.16$.  This per-galaxy accuracy is sufficient for
lensing work, because it is significantly less than the inherent shape
noise in each galaxy.  This is reflected in the design of the DLS,
which emphasizes area coverage and depth rather than an extended
filter set.  Similar results are obtained with a much larger
spectroscopic sample in a separate DLS field (Margoniner {\it et al.}, in
preparation).

In this subfield, the mean value of $\delta z$ averaged over all
redshifts is consistent with zero ($-0.0014 \pm 0.0077$), indicating
negligible bias.  However, the mean value obscures a tendency to
overpredict very low redshifts and underpredict redshifts near that of
the cluster.  For example, the mean $z_{\rm phot}$ of cluster members
is 0.60, as compared to the spectroscopic value of 0.68 which will be
derived in Section~\ref{sec-keck}.  We do not attempt to correct for
this trend here, as part of our purpose is to demonstrate how well
weak lensing tomography will work over very large areas without
spectroscopic feedback.  Due to the breadth of the lensing kernel, an
error of this size is still less than the statistical error in
locating a lens along the line of sight.

Photometric redshifts derived from this filter set are expected to
degrade for $z > 1.6$, because the 4000 \AA\ break is shifted through
the $\z$ filter.  We have no spectroscopic data to confirm such high
redshifts in this field, so we limit our analysis to sources with
$z_{\rm phot} < 1.6$.

\section{Weak lensing detection}

We measured weighted moments of objects in $R$ using the {\tt ellipto}
software described in Bernstein \& Jarvis (2002), discarding any
sources which triggered error flags.  We also used their seeing
correction procedure, discarding sources which were not at least 25\%
larger than the PSF.  We further winnowed the sources by requiring a
maximum observed ellipticity of 0.5 (rejecting about $15\%$ of
sources), because with $\sim$ 1\arcs\ resolution, highly elliptical
objects are quite likely to be blends of two distinct sources, based
on measurements of the Hubble Deep Field and synthetic fields
convolved with this seeing.  The number of sources passing all these
quality checks is 45435, or 37 arcmin$^{-2}$.

\begin{figure}
\epsscale{1.0}
\plotone{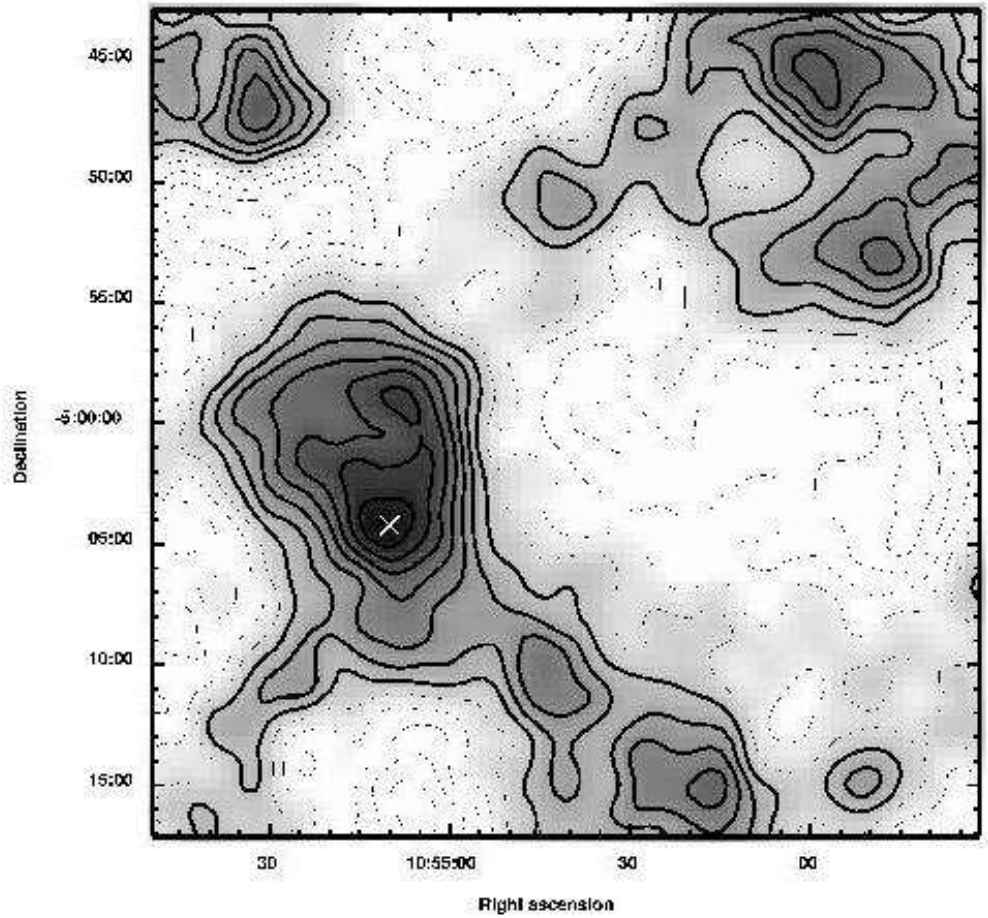}  
\caption{Projected mass map of the 35\arcm\ field, smoothed with a
30\arcs\ rms Gaussian.  Black indicates the most dense regions (the
units are arbitrary).  Contours are equally spaced from the lowest to
the highest value; negative and zero contours are drawn more thinly
than positive contours.  Note that only departures from the mean
density are measured, so that negative contours represent
underdensities.  The main mass concentration is coincident with a
cluster in which the brightest galaxy (location marked with an X) is
only 23\arcs\ from the peak projected mass density.
\label{fig-massmap}}
\end{figure}

We discovered the cluster before we had photometric redshift
information, by making a convergence map from $R$-selected sources,
using the method of Fischer \& Tyson (1997).  Figure~\ref{fig-massmap}
shows a convergence map made from selecting the 17163 sources with
$23<R<25$ from the final catalog, with 30\arcs\ pixels and smoothed
with a 30\arcs\ rms Gaussian.  A dominant mass concentration appears
in the figure, peaking at 10:55:11.6, -05:04:16 (all coordinates in
this paper are J2000).  Maps made from redshift-selected catalogs
appear similar to this one.  We show the $R$-selected map to
demonstrate that the weak lensing detection of this cluster does not
depend on photometric redshifts in any way.  We defer a discussion of
the significance of the detection to Section~\ref{sec-tomo}, where we
will take full advantage of the redshift information.

\begin{figure}
\epsscale{0.5}
\plotone{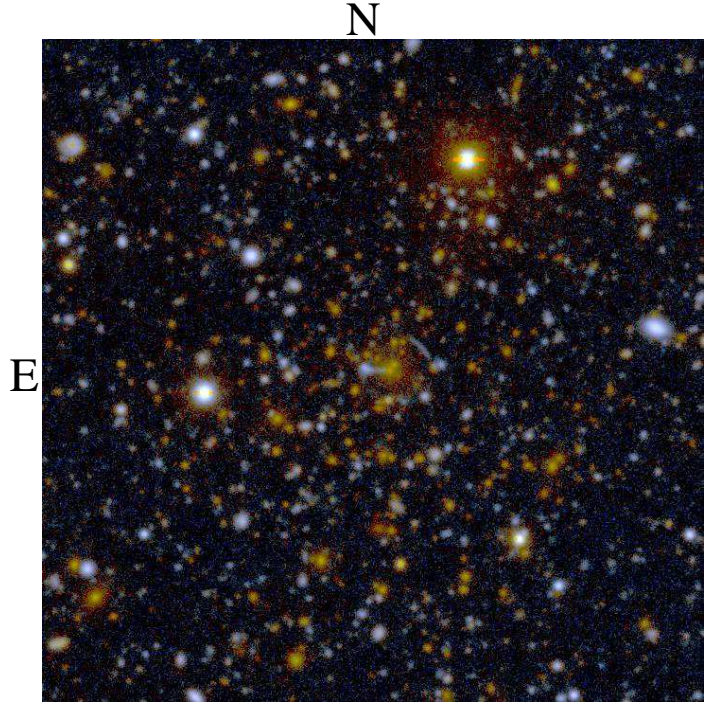}  
\caption{A 3\arcm\ square section of the $R$ image (of a BVR color
composite in the electronic edition),
centered on the BCG.  The BCG is $R=20.6$, and the faintest galaxies
visible in this reproduction are $R \sim 26$.  North is up,
and east to the left. The possible strong lensing arc is
10\arcs\ to the northwest of the brightest cluster galaxy.}
\label{fig-R}
\end{figure}

The multicolor imaging shows a concentration of red galaxies near the
location of the mass peak, with the brightest cluster galaxy (BCG) at
10:55:10.1 -05:04:13.  Figure~\ref{fig-R} shows a 3\arcm\ square
section of the $R$ image (of a $BVR$ color composite in the electronic
edition), centered on the BCG.  The BCG is 23\arcs\ from the mass
peak, well within the 1$\sigma$ mass peak positional uncertainty of
58\arcs\ derived from bootstrap resampling and from mass maps made
from a variety of similar, but statistically independent subcatalogs
representing different photometric redshift ranges.  Therefore we
tentatively identify the cluster with the mass concentration.  Neither
the cluster nor any apparent members are listed in the NASA/IPAC
Extragalactic Database or in the ROSAT All-Sky Survey Source Catalog.

Ten arcseconds to the northwest of the brightest cluster galaxy
appears an arc, which based on morphology alone is likely to be a
strongly lensed background galaxy.  Its redshift is unknown, but it is
bluer than the cluster members, consistent with the lensing
hypothesis.

\section{Spectroscopic Confirmation}
\label{sec-keck}

We took spectra of 24 likely member galaxies (with a projected
position near the cluster and in the magnitude range $20.7 < R <
22.6$) with the Low-Resolution Imaging Spectrograph (LRIS, Oke \etal\
1995) at W.~M. Keck Observatory in November, 2000.  Positions, $R$
magnitudes, redshifts, and spectral type and quality (following the
system of Cohen \etal\ 1999) are listed in Table~\ref{tab-z}.
Seventeen had redshifts in the range 0.664--0.694, representing a
cluster with a mean redshift of 0.68.  The remaining seven are
foreground and background galaxies in the redshift range 0.36--0.98.
Among the cluster members, the line-of-sight velocity dispersion is
$980 \pm 240$ km s$^{-1}$ using the biweight estimator of Beers \etal\
(1990).  The effect of membership uncertainty is modest: elimination
of the most deviant galaxy results in a biweight estimate of 840 km
s$^{-1}$.  We stress that the spectroscopy is used only as
confirmation, not as input to the photometric redshift and lensing
procedures.

The density of points in Figure~\ref{fig-zphot} seems to reveal a
second cluster, at $z=0.08$.  However, the galaxies at that
spectroscopic redshift are spread over the entire field and do not
appear to form a coherent cluster or group.  Furthermore, we find no
evidence for such a cluster in the lensing analysis below.  In any
case, the sensitivity of the lensing analysis to such a low-redshift
cluster would be low.  Whether this feature is an artifact of the
2dFGRS target selection procedure, an extremely diffuse group, or the
outskirts of a cluster outside the field, we conclude that it does not
affect the lensing analysis.

\section{Weak Lensing Tomography}
\label{sec-tomo}

As in W2001, we summarize the tangential shear $\gamma_t$ due to the lensing
cluster with a single number for each source redshift bin.  We do this
by separating the catalog into a series of source redshift slices,
then for each slice we compute the tangential shear for a series of
annuli centered on the BCG, fit a singular isothermal sphere (SIS)
profile to that data, and take the value of this fit (and its
uncertainty) at 1 Mpc projected radius.

\begin{figure}
\plotone{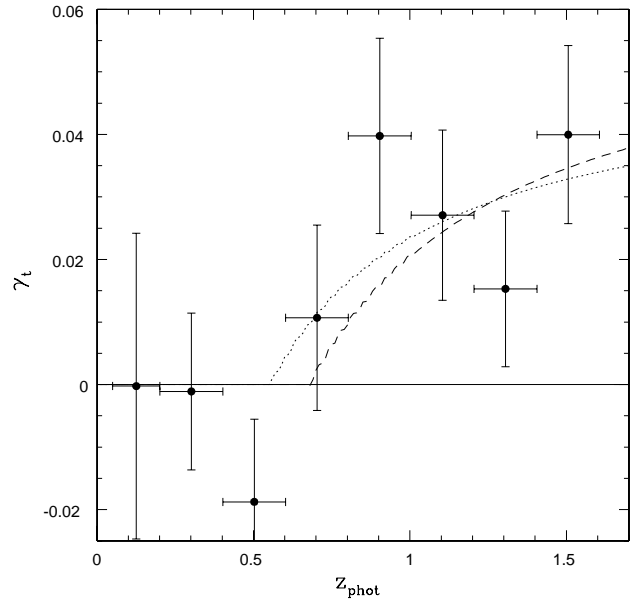}  
\caption{Tangential shear $\gamma_t$, centered on the BCG, as a function of
source photometric redshift.  The dotted curve shows the best-fit lens
fixed at the spectroscopic redshift of 0.68, and the dashed curve shows the
best fit when the lens redshift is allowed to vary ($z=0.55$).
}
\label{fig-tomo}
\end{figure}

\begin{figure}
\plotone{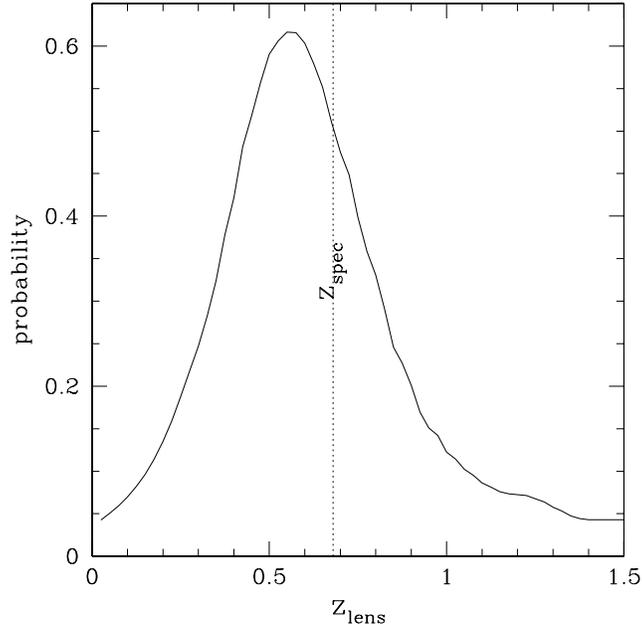}  
\caption{Lens redshift probability distribution.  The peak is at
$z=0.55$, the mean is $z=0.64$, and the rms is 0.29.  The
spectroscopic redshift of the cluster is marked with a vertical line
at $z=0.68$.  The agreement of the lens redshift with the
spectroscopic value indicates that any dark mass concentrations which
might be found in the DLS can be assigned rough redshifts (and
therefore masses and derived quantities such as M/L) from the lensing
information alone.
\label{fig-zprob}}
\end{figure}

Figure~\ref{fig-tomo} shows $\gamma_t$ as a function of source
photometric redshift.  The results are consistent with a lens at $z =
0.68$ (dashed line); the best-fit lens redshift is 0.55 (dotted line).
The full lens redshift probability distribution is shown in
Figure~\ref{fig-zprob}.  The mean and rms of this distribution are
0.64 and 0.29 respectively.  We also performed several null tests.  In
the first, we rotated each source by 45$^\circ$ and repeated the
analysis.  In the second null test, we repeated the analysis about
random centers.  In all these cases, the lens redshift probability
distribution is flat, with the best-fit lens at any redshift having
zero mass.  In contrast, under the zero-mass hypothesis the $\chi^2$
of the data in Figure~\ref{fig-tomo} is 22.4 for 7 degrees of freedom,
implying a probability of 0.2\%.  This is the best estimate of the
statistical significance of the discovery, taking advantage of both
redshift and shear information: We have 99.8\% confidence that these
data would not have arisen without a real lens.

Similar results, within the errors given, are obtained when using NFW
(Navarro, Frenk \& White 1997) profiles, changing the annular binning
scheme (by default, three logarithmically-spaced bins from 50\arcs\ to
30\arcm), varying the center by $\pm$1 arcminute, or changing the
redshift binning scheme provided the sampling is adequate.  We use
fixed-width redshift bins 0.2 wide, which gives a variable number of
sources per bin (from 653 to 4490 sources, increasing with redshift),
but maintains good redshift sampling.  

Note that we have neglected redshift bins above 1.6 due to the
limitations of the filter set.  We must also guard against sources at
$z>1.6$ contaminating the lower-redshift bins.  This cannot be done in
all generality, but with a lens at known redshift $z_{\rm lens}$, it
can be done for $z<z_{\rm lens}$.  There, a high rate of contamination
by high-redshift sources would increase the shear above its natural
value of zero.  From the low shear values observed for $z<0.6$ in
Figure~\ref{fig-tomo}, it would seem that such contamination is not a
major factor in the current dataset.  Given the large error bars, it
is difficult to put a precise limit on the contamination in the
current dataset (though we note that the very large error bars in the
lowest redshift bin reflect the paucity of sources, itself an
indication that high-redshift contamination is limited).  For the DLS
data as a whole, we will be able to derive limits using a sample of
clusters at a variety of redshifts.

\section{Mass estimates}

A first, rough mass estimate comes from strong lensing.  The arc
appears at a projected distance of 71 kpc.  If this is the Einstein
radius, the mass enclosed is $(1.8 \times 10^{13} \ M_\odot) {D_{s}
\over D_{ls}}$, where $D_s$ and $D_{ls}$ (in Gpc) are the angular
diameter distances from observer to source and from lens to source,
respectively.  ${D_{s} \over D_{ls}}$ could vary from unity (for
infinite source redshift) to perhaps five (for a source redshift of
0.9, which is a practical lower limit because the source is unlikely
to be in the small volume just behind the cluster).

To compare weak lensing and dynamical measurements on an equal
footing, we must adopt a model mass profile.  Following W2001 we adopt
a singular isothermal sphere (SIS) for its simplicity, as an NFW
profile requires an additional parameter but does not significantly
improve the fit to the shear profile (see below).  The velocity
dispersion then implies a projected mass of $7.0 \pm 3.4 \times
10^{14}\ (r/Mpc)\ M_\odot$ within radius $r$, or $5.0 \pm 2.4 \times
10^{13}\ M_\odot$ within 71 kpc, consistent with the strong
lensing estimate (all errors quoted are 1$\sigma$).

We estimate the mass using the weak lensing data in two different
ways, each time assuming $z_{\rm lens} = 0.68$.  First, we simply fit
the $\gamma_t(z_{\rm phot})$ data in Figure~\ref{fig-tomo} for the
lens mass, fixing $z_{\rm lens}$ at $0.68$.  The result is $11.1 \pm
2.8 \times 10^{14}\ (r/Mpc)\ M_\odot$ within radius $r$, or $7.9 \pm
2.0 \times 10^{13}\ M_\odot$ within 71 kpc, consistent with both
strong lensing and dynamical estimates.  Equivalently, the velocity
dispersion inferred from the lensing data is $1233 \pm 155$ km
s$^{-1}$.  Note that using the tomographic lens redshift of 0.55 in
the same formalism yields a $\sim 1 \sigma$ change in the mass
estimate, to $8.7 \pm 2.1 \times 10^{14}\ (r/Mpc)\ M_\odot$.

\begin{figure}
\plotone{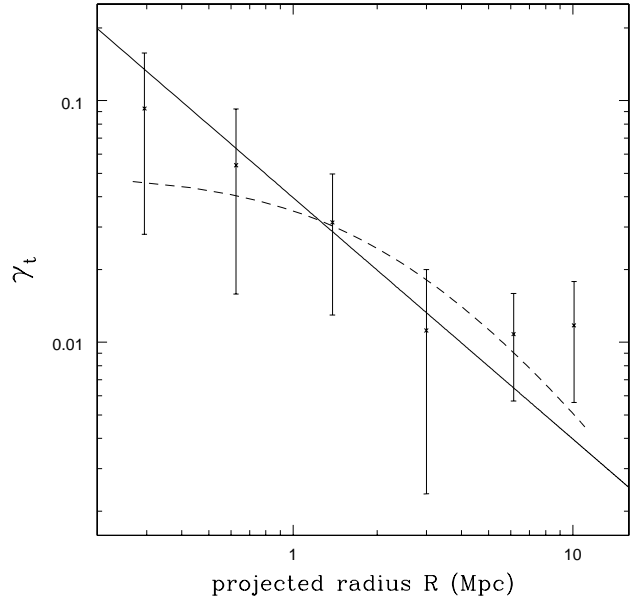}  
\caption{Radial shear profile of the cluster, using all sources with
$0.8 < z_{\rm phot} < 1.6$, along with best-fit SIS (solid line) and
NFW (dashed line) models.  The SIS fit is slightly better in terms of
$\chi^2$ per degree of freedom (0.58 with 5 degrees of freedom versus
0.65 with 4 degrees of freedom for the NFW).  Because the SIS model
fits the data better with one fewer parameter, it is used throughout
this paper.  The SIS model shown contains $8.6 \pm 2.3 \times 10^{14}
\ (r/Mpc)\ M_\odot$ within radius $r$.
\label{fig-radial}}
\end{figure}

Alternatively, we attempt to constrain the radial profile more
strongly by making a single radial profile using all sources at $0.8 <
z_{\rm phot} < 1.6$, which is more like a traditional weak lensing
analysis with a simple foreground/background cut.  This radial profile
is shown in Figure~\ref{fig-radial}.  A straightforward SIS fit to
these data (solid line in the figure) yields $8.6 \pm 2.3 \times
10^{14}\ (r/Mpc)\ M_\odot$ within radius $r$, consistent with all the
other estimates.  The inferred velocity dispersion is $1085 \pm 128$
km s$^{-1}$.  An NFW fit is also shown (dashed line).  The SIS fit is
slightly better in terms of $\chi^2$ per degree of freedom (0.58 with
5 degrees of freedom versus 0.65 with 4 degrees of freedom for the
NFW).  Given that both are acceptable fits, use of the simpler
one-parameter SIS model throughout this paper is justified.

Finally, we estimate the mass-to-light (M/L) ratio using the first
weak lensing mass quoted above.  We measured the light in $z'$ band,
which roughly corresponds to emitted $V$ band.  This time we made
catalogs in single-image mode on the $z'$ image so as not to miss very
red sources.  We extracted subcatalogs centered on the cluster and on
seven control regions, and computed the total magnitude of sources
within those regions meeting two criteria: being resolved (eliminating
stars), and with a $z'$ magnitude between 19.75 (the magnitude of the
BCG) and 23.75, beyond which incompleteness starts to set in.  The
total measured magnitude of the cluster (within a radius of 500 kpc)
minus the mean background is $z' = 15.86 \pm 0.18$ mag (the
uncertainty comes from variation among the control regions).  After
applying a small correction for the faint end of the luminosity
function which was missed and converting from observed $z'$ band to
rest-frame $V$ using the approach of Fischer \& Tyson (1997), we find
a rest-frame $M/L_V = 574 \pm 146$.  This value is quoted at a
projected radius of 500 kpc, but it does not vary significantly in the
projected radius range 250---1000 kpc.

\section{Conclusions}

We have extended shear selection of clusters to a higher redshift
range, which will be cosmologically useful.  For example, since dark
energy has its largest effect on comoving volume at redshift
$\sim$0.5, measurements of the volume-redshift relation via mass
cluster counting must bracket this and extend up to $z\sim 1$ (Tyson
\etal\ 2003).  Furthermore, using the observed $\gamma_t(z_{\rm
phot})$ relations, we have identified the redshift of the {\it lens}
in addition to that of the cluster, and found that they are
consistent.  Thus any dark clusters which might be found in the DLS
can be assigned rough redshifts and masses, a necessary first step in
investigating them, as well as including them in cluster-counting
cosmological tests.

The mass of the cluster is fairly high, but by no means exceptional.
Some examples of more massive clusters at this or higher redshift are
MS1054-03 at $z=0.83$, with a velocity dispersion of $\sigma_v = 1170$
km s$^{-1}$ (Tran \etal\ 1999) and $\sigma_v = 1311$ km s$^{-1}$ inferred
from weak lensing (Hoekstra \etal\ 2000), and CL1604+4304 at $z=0.90$,
with $\sigma_v \sim 1200$ km s$^{-1}$ (Postman, Lubin \& Oke 2001).
This suggests that even at high redshift, the DLS is sensitive to
clusters over a significant range of the mass function.  However, any
conclusions as to the nature of the overall DLS sample would be
entirely premature.  This cluster may not be representative, as the
large arc was a significant factor in choosing to investigate this
cluster first.

The M/L of this cluster is very high, but examples of darker clusters
can be found.  For example, Fischer (1999) found $M/L_R = 640 \pm 150$
for MS12247+2007, consistent with the earlier measurement of Fahlman
\etal\ (1994) on the same cluster.  Whether shear-selected clusters
tend to be {\it systematically} underluminous (or more accurately,
whether optically-selected clusters tend to be overluminous)
is a fascinating question which awaits the compilation of a
statistically significant sample.  We note that the M/L of the ``dark
clump'' detected by Erben \etal\ could be as low as $\sim 400$,
depending on its redshift which is still unknown (Gray \etal 2001).
Thus the label ``dark'' may well be misleading if it implies a new
class of objects.  There may well be a continuous distribution
encompassing all these examples as well as optically-selected
clusters.

A difficulty with optical M/L ratios is that they depend greatly on
star formation history, which tends to obscure the underlying question
of how mass is assembled.  These astrophysical processes are more
directly related to the X-ray properties of clusters, so an even more
fascinating question which awaits the compilation of a statistically
significant sample is whether shear-selected clusters tend to be X-ray
underluminous.  This and other shear-selected clusters from the DLS
are currently being followed up with {\it Chandra} and {\it XMM} X-ray
imaging.

A cluster of this mass is not unexpected in the volume probed by these
data (Rahman \& Shandarin 2001), so the number of clusters in the
complete DLS could be estimated by scaling up the area sampled here,
yielding $\sim$60.  However, that is a lower limit because we chose
only the single most dominant mass concentration in this area.
Preliminary analysis of full-depth areas suggests that about 200
clusters will be found in the DLS.  The tightness of constraints on
cosmological parameters afforded by a sample of this size, with and
without priors from other measurements such as the cosmic microwave
background, are being computed (Hennawi \& Spergel, in preparation).

\acknowledgments 

We thank NOAO for supporting survey programs, and the 2dFGRS project
for making data publicly available.  Observations were obtained at
Cerro Tololo Inter-American Observatory and the W. M. Keck
Observatory.  CTIO is a division of National Optical Astronomy
Observatory (NOAO), which is operated by the Association of
Universities for Research in Astronomy, Inc., under cooperative
agreement with the National Science Foundation.  The W.M. Keck
Observatory is operated jointly by the California Institute of
Technology, the University of California, and the National Aeronautics
and Space Administration.  This work also made use of IRAF and of the
NASA/IPAC Extragalactic Database.

\clearpage

\begin{table}
\caption{Spectroscopic Redshifts}
\label{tab-z}
\begin{tabular}{|llccll|}
\tableline 
RA (J2000) & DEC & m$_R$ & Type\tablenotemark{a} & Quality\tablenotemark{a} & z \\
\tableline
\multicolumn{6}{|c|}{Cluster members} \\
\tableline
10:55:15.4 & -05:05:14 & 20.57 & AI & 1 & 0.680\\
10:55:09.1 & -05:06:22 & 21.15 & A & 1 & 0.680\\
10:55:11.8 & -05:05:31 & 20.89 & A & 1 & 0.678\\
10:55:12.1 & -05:04:35 & 22.23 & A & 1 & 0.688\\
10:55:11.1 & -05:04:37 & 22.24 & A & 2 & 0.694\\
10:55:12.3 & -05:04:09 & 21.35 & AI & 1 & 0.679\\
10:55:10.7 & -05:04:18 & 21.91 & A & 2 & 0.681\\
10:55:10.1 & -05:04:13 & 20.60 & A & 1 & 0.677\\
10:55:10.4 & -05:04:12 & 21.00 & A & 2 & 0.674\\
10:55:08.8 & -05:04:01 & 22.33 & A & 2 & 0.669\\
10:55:08.4 & -05:03:31 & 21.81 & A & 1 & 0.688\\
10:55:06.8 & -05:03:18 & 21.74 & A & 1 & 0.680\\
10:55:02.6 & -05:02:03 & 21.21 & I & 1 & 0.675\\
10:54:57.8 & -05:01:50 & 21.81 & I & 1 & 0.664\\
10:54:55.0 & -05:01:07 & 20.80 & I & 1 & 0.670\\
\tableline
\multicolumn{6}{|c|} {Non-members} \\
\tableline
10:55:07.2 & -05:03:23 & 20.20 & A & 1 & 0.364\\
10:55:04.8 & -05:02:31 & 20.72 & I & 1 & 0.384\\
10:55:19.1 & -05:04:49 & 21.24 & I & 1 & 0.523\\
10:55:12.8 & -05:04:54 & 20.93 & EI & 1 & 0.720\\
10:55:05.6 & -05:02:53 & 21.53 & A & 1 & 0.728\\
10:54:55.8 & -05:00:41 & 21.71 & A & 1 & 0.731\\
10:54:59.7 & -05:00:52 & 21.73 & E & 4 & 0.984\\      
\tableline
\end{tabular}
\tablenotetext{a}{Type and quality follow the system of Cohen \etal
(1999), in which ``A'' indicates absorption-dominated, ``E'' indicates
emission-dominated, and ``I'' indicates an intermediate spectrum.}
\end{table}

\clearpage

\end{document}